\documentclass{ametsocV6.1}

\usepackage{threeparttable}
\usepackage{multirow}
\usepackage{makecell}

\title{A Reduced-Order Coupled Oscillator Model for the Emergence of Episodic Convection}

\authors{
Sooman Han,\aff{a}\correspondingauthor{Sooman Han, sooman.han@yale.edu}
Soong-Ki Kim,\aff{b}
Bowen Fan,\aff{a}
Jérôme Vialard,\aff{c}
Alexey V. Fedorov,\aff{a,c}
and Juan M. Lora\aff{a}
}

\affiliation{
\aff{a}{Department of Earth and Planetary Sciences, Yale University, New Haven, CT, USA}\\
\aff{b}{Division of Environmental Science and Engineering, Pohang University of Science and Technology, Pohang, South Korea}\\
\aff{c}{LOCEAN-IPSL, Sorbonne Université -CNRS-IRD-MNHN, Paris, France}
}

\abstract{Extreme precipitation is a societal concern, and understanding tropical moist convection is increasingly important in a warming climate. Climate simulations show that, under hothouse Earth, convection can transition from quasi-steady precipitation to an episodic regime characterized by intense rainfall bursts separated by extended dry intervals, although the mechanisms governing this transition remain debated. Here we develop a reduced-order framework that represents tropical convection through column-integrated moisture ($q$) and thermal stratification ($\Delta$), rather than bulk measures of convective instability and inhibition. Analysis of cloud-resolving model (CRM) simulations shows that these variables evolve together: during dry periods, surface fluxes recharge moisture, while radiative and dynamical processes erode the stratification generated by previous convection. We couple this evolution to the observed nonlinear moisture dependence of tropical precipitation ($P$), whereby rainfall increases sharply above a critical moisture threshold. A third prognostic variable, $C$, represents convective persistence, allowing established convection to continue below its onset threshold and thereby producing hysteresis. With parameters constrained by CRM output, the model closely reproduces the simulated evolution of $q$, $\Delta$, and $P$ in both convective regimes, representing quasi-steady convection as stable, noise-driven oscillations and episodic convection as an unstable, self-sustained limit cycle. Linear stability analysis shows that the transition depends jointly on moisture-recharge and stratification-adjustment timescales, thermal–moisture sensitivity, and convective persistence. Because radiative and dynamical processes enter through their effects on thermal stratification, the framework is not tied to a specific mechanism, providing a unified dynamical link between observed moisture-threshold behavior of precipitation and emergence of episodic convection under warming.
}

\begin{document}

\maketitle

%
%
%
%
%
%

\statement
Precipitation sustains societies and ecosystems, but extreme downpours and prolonged dry periods can cause severe damage. Climate simulations suggest that under hothouse conditions resembling Earth’s distant past or far future, with tropical sea-surface temperatures above approximately 320–325 K, precipitation may occur in intense episodes separated by long dry intervals. In contrast, present-day precipitation is generally weaker and more frequent. Observations also show that tropical precipitation increases sharply above a critical atmospheric-moisture threshold. We develop a simple model connecting these behaviors. It shows how interactions between atmospheric moisture and thermal structure produce either weak, steady precipitation or recurring episodic bursts, depending on moisture-recharge and thermal-adjustment timescales and thermal–moisture sensitivity strength. This framework isolates the essential mechanisms governing transitions between precipitation regimes.

%

\section{Introduction}

Precipitation sustains freshwater resources, agriculture, and ecosystems, yet extreme rainfall can cause destructive flooding, making precipitation variability a major societal concern \citep{Seneviratne21, Caretta22}. In the tropics, precipitation is tightly coupled to atmospheric moisture and moist convection \citep{Emanuel94, Bretherton04, Schneider10}, and its extremes are particularly sensitive to warming \citep{Gorman09, Muller11, Romps11, Gorman12, Gorman15, Neelin22}. Therefore, understanding the processes that regulate precipitation is essential for improving predictions of its variability and extremes in a warming climate \citep{IPCC23}.

Recent studies have found that convection transitions from a quasi-steady to an episodic regime in hothouse climates with sea surface temperatures (SSTs) exceeding approximately 320--325~K. Quasi-steady convection produces continuous, weak precipitation of \(\mathcal{O}(1)\)~mm day\(^{-1}\), whereas episodic convection produces intense rainfall bursts of \(\mathcal{O}(100)\)~mm day\(^{-1}\) separated by multi-day dry spells \citep[e.g.,][]{Seeley21,Dagan23,Liu23,Spaulding24,Habib26}. This transition was initially attributed to lower-tropospheric radiative heating (LTRH). In warm, moist atmospheres, closure of infrared windows and increased absorption of solar radiation suppress surface-based convection, allowing moist static energy to accumulate until evaporative cooling by virga erodes the inhibition and triggers a deluge \citep{Seeley21}. Subsequent studies showed that positive LTRH is sufficient but not necessary: episodic convection can also emerge when radiative cooling is stronger in the upper troposphere than below \citep{Dagan23,Song24}. These results indicate that multiple radiative heating and cooling patterns can generate the necessary convective inhibition and release.

Importantly, episodic convection is not limited to extreme hothouse conditions above 320--325~K. Absorbing aerosols can generate LTRH and trigger episodic convection at SSTs of 305--310~K \citep{Dagan24,Sreelekshmi26}. General circulation and planetary-scale cloud-resolving simulations further show that convectively coupled gravity waves can organize periodic extreme precipitation over the same temperature range \citep{Quan26}. Episodic convection may therefore emerge under less extreme warming, or through aerosol and circulation changes, making it potentially relevant to regional precipitation extremes in a changing climate rather than only to a distant hothouse state.

Separately, satellite observations of the tropics have revealed a sharp, nonlinear increase in precipitation beyond a moisture-dependent convective threshold \citep[e.g.,][]{Peters06, Neelin09, Holloway09, Holloway10, Kuo17, Kuo18}. Once atmospheric moisture exceeds this threshold, precipitation rates rise sharply to $\mathcal{O}(100)\ \mathrm{mm\ day^{-1}}$ from $\mathcal{O}(1)\ \mathrm{mm\ day^{-1}}$, a regime termed \textit{strong convection} \citep{Neelin09}. Previous conceptual precipitation models expressed the nonlinear onset of strong convection by introducing a threshold-like dependence on atmospheric moisture \citep{Muller09, Stechmann11, Stechmann14}. Here, it is important to distinguish \textit{strong convection} from \textit{episodic convection}. The former refers to the sharp increase in precipitation above a critical moisture threshold observed under present-day tropical conditions, which describes a state-dependent moisture--precipitation relationship. The latter refers to an oscillatory dynamical regime characterized by intense precipitation events separated by extended dry intervals, which can emerge through changes in SST and/or radiative heating. Although both behaviors reflect nonlinearity in precipitation, whether the moisture threshold associated with strong convection contributes to the emergence of episodic convection remains unexplored.

Given the growing number of climate model simulations exhibiting transitions between quasi-steady and episodic convection, a couple of conceptual theories have previously been developed to explain the transition. By combining a zero-buoyancy bulk-plume model with a heat-engine model, \citet{Spaulding24} proposed that this transition occurs when convective available potential energy (CAPE) is insufficient to support the convective mass transport required by the heat engine. \citet{Yang24} formulated a predator--prey model using precipitation and convective inhibition (CIN) as prognostic variables, where increasing the prescribed mean CIN causes the equilibrium to lose stability through a supercritical Hopf bifurcation, producing self-sustained precipitation oscillations. Although these frameworks provide important explanations for the onset of episodic convection, they are formulated in terms of CAPE, CIN, and precipitation, which are bulk diagnostics that do not isolate the respective contributions of atmospheric moisture and thermal structure. Consequently, the distinct roles of moisture and thermal structure in controlling the regime transition remain unclear. Moreover, neither framework explicitly connects the moisture-threshold behavior associated with \textit{strong} convection to the emergence of \textit{episodic} convection.

Given this background, we develop a zero-dimensional conceptual model that couples threshold-like moisture--precipitation behavior to atmospheric thermal structure, thereby linking strong and episodic convection. Our highly reduced conceptual model explicitly couples column-integrated atmospheric moisture content and thermal stratification as its two primary thermodynamic prognostic variables, and is not tied to any specific radiative or dynamical mechanism governing the convective regime. Rather than conducting additional climate simulations, we use the cloud-resolving model (CRM) simulations of \citet{Seeley21} as a reference and evaluate our conceptual model against their results. In Section~2, we characterize the quasi-steady and episodic convective regimes in the CRM and identify the coupled behavior between thermal stratification and atmospheric moisture. In Section~3, we develop our conceptual model using these quantities as prognostic variables. In Section~4, we evaluate our model against the CRM simulations and demonstrate that it reproduces the principal characteristics of both convective regimes. We discuss the physical interpretation of the transition between these regimes and conclude in Section~5.

\section{Cloud Resolving Model}

We analyze archived output from the CRM simulations of \citet{Seeley21}. Their baseline simulations employ the non-hydrostatic Das Atmosphärische Modell (DAM; \citealt{Romps08}) over a doubly periodic $72\times72~\mathrm{km}^2$ domain with 2-km horizontal grid spacing and prescribed SST. The model has 140 vertical levels, with grid spacing increasing from $\Delta z=25~\mathrm{m}$ below 650~m to 500~m between 5.4 and 33~km and 1,000~m above 38~km. The archived data consist of domain-mean output spanning 200 days with an hourly time step. We use simulations with SSTs of 305 and 325~K as representative of the quasi-steady and episodic convection regimes, respectively. We refer readers to \citet{Seeley21} for further details of their CRM configuration.

Figure~1 shows vertical profiles of radiative heating, potential temperature, and specific humidity from the CRM simulations with SSTs of 305 and 325~K, averaged over the final 20~days of each simulation. Figures~1(a,b) show that the 305~K simulation exhibits radiative cooling throughout the troposphere except within a thin surface-adjacent heating layer, whereas the 325~K simulation develops LTRH below approximately 4--5~km, except within a thin surface-adjacent cooling layer, overlain by radiative cooling aloft. \citet{Seeley21} did not diagnose the origin of these opposite-signed tendencies in the lowermost model layer; they may reflect a warming-induced change in radiative flux divergence near the lower boundary, as increased water-vapor opacity and shortwave absorption alter the longwave exchange between the prescribed-temperature surface and the surface-adjacent atmosphere. A detailed investigation of this feature is beyond the scope of our study. The LTRH has been invoked to explain episodic convection because it suppresses the ascent of near-surface air during the dry phase and stabilizes the lower troposphere. Figures~1(c,d) show that the vertical gradient of potential temperature strengthens with warming. This increased thermal stratification follows from moist-adiabatic thermodynamics: as the atmosphere warms, saturated ascending air contains more water vapor and releases more latent heat. This additional heating reduces the rate at which temperature decreases with height, causing potential temperature to increase more rapidly with height and producing a more stably stratified troposphere \citep{Quan26}. Finally, Figures~1(e,f) show that specific humidity increases throughout the atmospheric column with warming, consistent with Clausius--Clapeyron scaling. In other words, increased thermal stratification and atmospheric water vapor content are defining features of the warmer climate state.

\begin{figure}[htbp]
  \centering
  \includegraphics[width=0.85\textwidth]{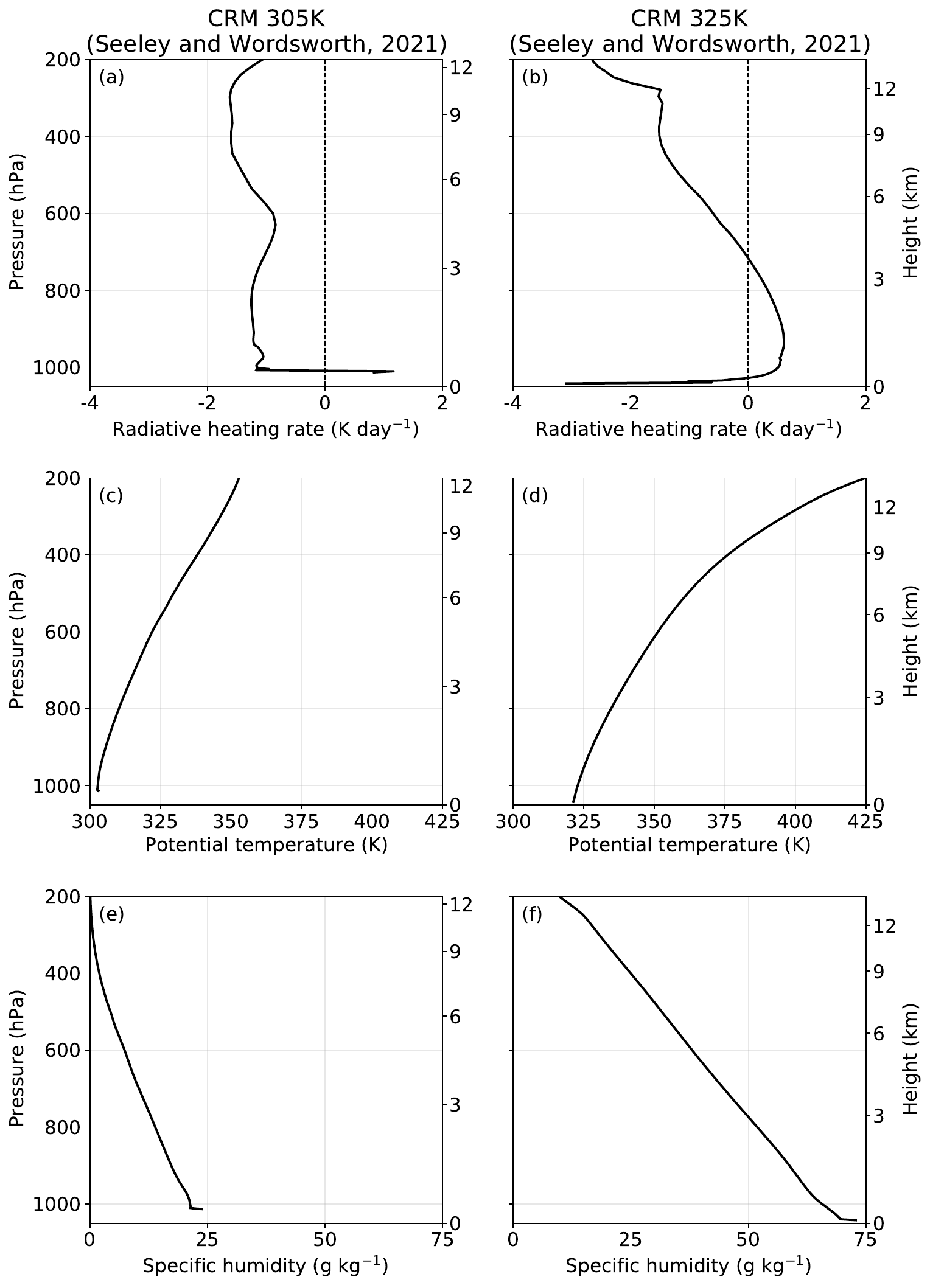}
  \caption{Vertical profiles of radiative heating (panels a,b), potential temperature (panels c,d), and specific humidity (panels e,f) from the cloud-resolving model simulations of \citet{Seeley21}. The left and right columns correspond to SSTs of 305 and 325~K, respectively. All profiles are averaged over the final 20-day analysis period of each simulation.}
\end{figure}

Figure~2 shows time series of precipitation, $P$ ($\mathrm{mm\ day^{-1}}$), thermal stratification $\Delta$ (K), and column-integrated moisture $q$ (mm), together with the corresponding $\Delta$--$q$ phase diagrams, for the same CRM simulations in Figure~1. Thermal stratification is defined as the difference between the potential temperatures of the middle and lower troposphere:
\begin{equation}
\Delta
\equiv
\left\langle
\theta
\right\rangle_{650\text{--}450\,\mathrm{hPa}}
-
\left\langle
\theta
\right\rangle_{1000\text{--}800\,\mathrm{hPa}},
\label{eq:thermal_stratification}
\end{equation}
where the angle brackets denote a pressure-weighted vertical mean. Figures~2(a,b) show quasi-steady precipitation at 305~K and episodic precipitation at 325~K. These regimes can be quantified by the regime index $\eta \equiv \frac{\sigma_P}{\overline{P}}$ where $\overline{P}$ and $\sigma_P$ denote the temporal mean and standard deviation of precipitation, respectively \citep{Dagan23}. The 305~K simulation has $\eta=0.25<1$, indicating quasi-steady precipitation, whereas the 325~K simulation has $\eta=5.08>1$, indicating episodic precipitation.

\begin{figure}[htbp]
  \centering
  \includegraphics[width=1.0\textwidth]{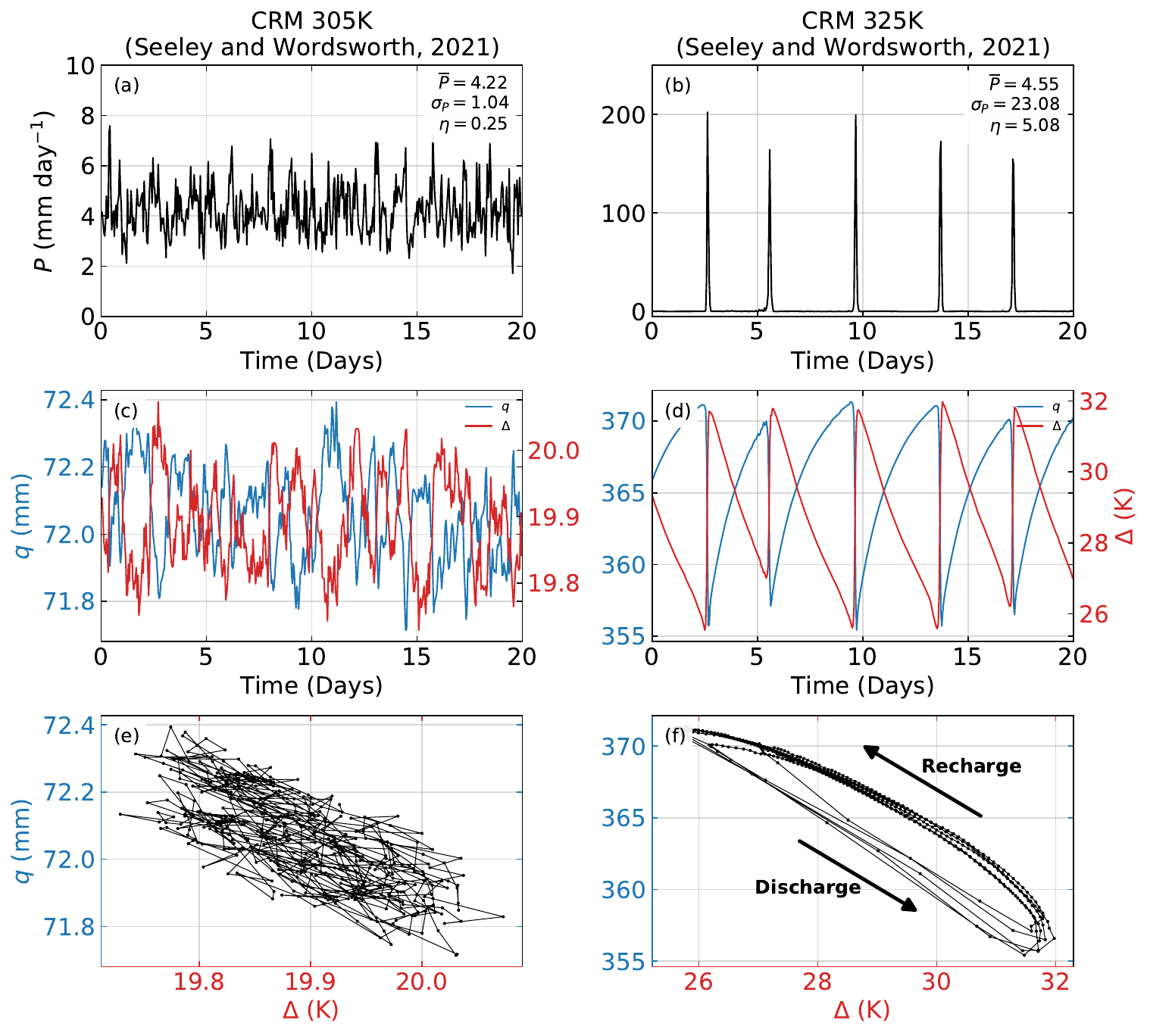}
  \caption{
  Time series of precipitation $P$ (mm day$^{-1}$; panels a,b) and thermal stratification $\Delta$~(K) and column-integrated moisture~$q$ (mm; panels~c,d), together with the corresponding $\Delta$--$q$ phase diagrams (panels~e,f), for the 305~K (left column) and 325~K (right column) SST CRM cases. Arrows in panel~(f) indicate the recharge and discharge phases of episodic convection. Thermal stratification is defined as $\Delta \equiv \left\langle\theta\right\rangle_{650\text{--}450\,\mathrm{hPa}}-\left\langle\theta\right\rangle_{1000\text{--}800\,\mathrm{hPa}}$, where the brackets denote pressure-weighted mean potential temperature. $\overline{P}$ and $\sigma_P$ denote the temporal mean and standard deviation of precipitation, respectively. Following \citet{Dagan23}, the precipitation-regime index is defined as $\eta \equiv \sigma_P/\overline{P}$, with $\eta<1$ indicating quasi-steady precipitation and $\eta>1$ indicating episodic precipitation.
  }
\end{figure}

Importantly, Figures~2(c,d) reveal coupled variations in $\Delta$ and $q$ in both regimes. Between convective events, surface evaporation replenishes column moisture, causing $q$ to increase. Meanwhile, differential radiative heating and cooling erode the stratification generated by previous convection, causing $\Delta$ to decrease \citep{Seeley21}. Together, these processes constitute the recharge phase. When convection develops, precipitation removes atmospheric moisture, while latent heating concentrated aloft rebuilds thermal stratification. This sequence produces the anticorrelated evolution of $\Delta$ and $q$. Their coupled evolution is irregular and dominated by noise at 305~K but becomes more regular and periodic at 325~K.

The $\Delta$--$q$ phase diagrams (Figures~2(e,f)) further distinguish the two regimes. At 305~K, the trajectory fluctuates irregularly without a persistent closed orbit, consistent with a damped oscillator continually excited by noise \citep{Strogatz18}. At 325~K, it forms a repeating closed orbit, or limit cycle, characteristic of a self-sustained oscillator. This contrast has a close analogue in recharge-oscillator dynamics of El Niño--Southern Oscillation (ENSO), which describe coupled variations in ocean heat content and the Niño index \citep[e.g.,][]{Vialard25}. \citet{Jin97} showed how ENSO can operate as an unstable, self-sustained recharge oscillator, whereas subsequent studies found that a stable, noise-driven oscillator provides a more realistic representation of observed ENSO variability \citep{Thompson01, Han26a, Han26b}. This analogy is particularly relevant because climate change may potentially shift ENSO toward a less damped or self-sustained regime \citep{Timmermann01, Stuecker25}. Our analysis suggests that an analogous transition may occur in tropical convection.

Additionally, the 325~K case exhibits clear hysteresis: precipitation begins near $q\approx370~\mathrm{mm}$ and persists until $q$ decreases to approximately $355~\mathrm{mm}$. Thus, at the same intermediate moisture content (e.g., $q\approx360~\mathrm{mm}$), the system may occupy either the slow recharge branch (upper right) or the rapid convective-discharge branch (lower left), depending on its prior state. This behavior is consistent with the stochastic hysteresis identified by \citet{Stechmann11}. In their two-state stochastic model, both precipitating and nonprecipitating states are possible at the same $q$ because a convective event initiated at high $q$ may persist as precipitation depletes moisture to values at which a new event would be unlikely to begin. In the 325~K phase diagram, the longer residence time during moisture recharge produces denser sampling along the recharge branch.

Figures~A1 and A2 repeat the analyses of Figures~1 and 2 using two additional CRM simulations from \citet{Song24}, an independent modeling study of episodic convection: their 305~K experiment and their 325~K polar-night experiment, respectively. This comparison tests whether the coupled evolution of $\Delta$ and $q$ depends on the particular radiative mechanism producing episodic convection. The 305~K setup is similar to its counterpart in \citet{Seeley21}, with predominantly radiative cooling throughout the troposphere, and likewise exhibits quasi-steady convection. In contrast, the 325~K polar-night setup eliminates shortwave heating, and the prescribed radiative profile removes LTRH while retaining substantially stronger cooling in the upper troposphere than in the lower troposphere; nevertheless, it produces episodic convection. Despite the distinct radiative profiles in the warm simulations of \citet{Seeley21} and \citet{Song24}, both exhibit the same anticorrelated evolution of $q$ and $\Delta$. The phase diagrams from the simulations of \citet{Song24} display a similar dynamical distinction: irregular fluctuations typical of noise-driven oscillations in a stable system at the lower SST, and a closed orbit typical of a self-sustained regime in the warmer polar-night experiment, with the latter again exhibiting hysteresis.

The occurrence of episodic convection in both studies, despite their distinct radiative heating properties, can be understood qualitatively as follows. In \citet{Seeley21}, LTRH accompanied by radiative cooling aloft warms the lower troposphere relative to the upper troposphere, thereby reducing thermal stratification. In \citet{Song24}, radiative cooling occurs throughout the troposphere, but stronger cooling aloft than below likewise reduces thermal stratification. In both 325~K cases, the warmer atmosphere is more strongly stratified, as expected from its warmer moist-adiabatic profile, so radiative processes require more time to erode this stability before deep convection can resume. Thus, both produce a slow reduction in $\Delta$ while moisture recharges. These results demonstrate that the coupled $\Delta$--$q$ dynamics are robust across distinct pathways to episodic convection that nevertheless share a common bulk evolution.

\section{Coupled Oscillator Model}

Motivated by the CRM analysis in the preceding section, we construct a reduced-order coupled oscillator model (COM) with three prognostic variables: column moisture $q$, thermal stratification $\Delta$, and convective persistence $C$. The variable $C$, described in detail below, is introduced specifically to capture the hysteresis seen in the CRM by allowing established convection to continue after the conditions governing its onset are no longer satisfied. Depending on the governing parameters, the model reproduces quasi-steady convection as stable, noise-driven oscillations and episodic convection as unstable, self-sustained oscillations with hysteresis.

\subsection{Model Formulation}

We begin with a threshold representation of convective precipitation
based on a hyperbolic tangent function \citep{Stechmann11}:
\begin{equation}
P
=
\frac{P_{\max}}{2}
\left[
1+\tanh(X)
\right],
\qquad
X
=
\frac{q-q_{\mathrm{crit}}}{q_{\mathrm{width}}}.
\label{eq:precipitation}
\end{equation}
Here, $P_{\max}$ is the upper bound on the convective precipitation rate $(\mathrm{mm\,day^{-1}}$); $q_{\mathrm{crit}}$ is the critical moisture content required for strong convective onset (mm); and $q_{\mathrm{width}}$ controls the width of the transition between weak and strong precipitation (mm). This functional form gives $P\approx0$ when $q\ll q_{\mathrm{crit}}$ and $P\approx P_{\max}$ when $q\gg q_{\mathrm{crit}}$, while $q_{\mathrm{width}}$ determines how sharply the transition occurs around $q=q_{\mathrm{crit}}$.

The precipitation function enters the moisture-budget equation, which is
\begin{equation}
\frac{\mathrm{d}q}{\mathrm{d}t}
=
\frac{q_0-q}{\tau_q}
-P+\sigma_q w_q.
\label{eq:moisture_budget}
\end{equation}
The first term on the right-hand side of
Eq.~\eqref{eq:moisture_budget} represents moisture recharge toward the reference value $q_0$ (mm) over the timescale $\tau_q$ (days) by surface evaporation, and the second term represents moisture removal by precipitation. The third term represents processes unresolved by the first two terms, such as moisture convergence and divergence, as stochastic perturbations \citep{Stechmann11, Hottovy15a, Hottovy15b}. Here, $w_q$ is normalized Gaussian white noise satisfying $\langle w_q(t)w_q(t')\rangle=\delta(t-t')$. Because the delta function has units of $\mathrm{day}^{-1}$, $w_q$ has units of $\mathrm{day}^{-0.5}$; therefore, $\sigma_q$ has units of $\mathrm{mm\ day}^{-0.5}$.

The thermal stratification equation is given as 
\begin{equation}
\frac{\mathrm{d}\Delta}{\mathrm{d}t}
=
\frac{\Delta_0-\Delta}{\tau_\Delta}
+\gamma P+\sigma_{\Delta}w_{\Delta}.
\label{eq:thermal_budget}
\end{equation}
The first term on the right-hand side of Eq.~\eqref{eq:thermal_budget} represents thermal relaxation toward the reference stratification $\Delta_0$ (K) over the timescale $\tau_\Delta$ (days). This term represents the erosion of stratification generated by convection, and we therefore interpret $\tau_\Delta$ as a stratification-adjustment timescale. LTRH combined with cooling aloft—or, more generally, stronger radiative cooling aloft than below—reduces $\Delta$ by cooling the upper layer relative to the lower layer. Dynamical adjustment associated with processes such as coupled gravity waves can likewise redistribute the thermal structure and weaken the stratification anomaly \citep[e.g.,][]{Quan26}. The second term represents convectively-induced stabilization of the troposphere, where the coefficient $\gamma$ has units of $\mathrm{K\ mm^{-1}}$. $\sigma_{\Delta}w_{\Delta}$ represents unresolved stochastic processes affecting thermal stratification, such as fluctuations in cloud feedback, convective entrainment and detrainment, and transient vertical advection associated with gravity waves. Here, $w_{\Delta}$ is normalized Gaussian white noise with units of $\mathrm{day}^{-0.5}$, and $\sigma_{\Delta}$ is its amplitude with units of $\mathrm{K\ day}^{-0.5}$.


We allow the moisture threshold for convection to vary with two additional aspects of the atmospheric state. First, the threshold depends on thermal stratification because a more stable thermal profile increases the buoyancy barrier faced by boundary-layer parcels. Second, established convection can persist at lower moisture levels than those required for its initial onset, representing the hysteresis seen in the CRM. We capture these effects through a baseline moisture threshold $q_{\mathrm{crit},0}$, a thermal sensitivity $\alpha$, and the convective-persistence variable $C$:

\begin{equation}
q_{\mathrm{crit}}
=
q_{\mathrm{crit},0}
-
\alpha\left(\Delta-\overline{\Delta}\right)
-
q_{\mathrm{hyst}}C,
\label{eq:qcrit}
\end{equation}

\noindent where $\overline{\Delta}$ denotes the temporal-mean stratification, treated as a fixed reference value. The parameter $q_{\mathrm{crit},0}$ is the baseline moisture threshold at $\Delta=\overline{\Delta}$ and $C=0$, while $\alpha$ controls the sensitivity of the threshold to thermal stratification. Within each climate simulation, $q_{\mathrm{crit},0}$ is a fixed parameter, fitted separately for the 305 and 325~K cases. We expect its value to increase with the background tropospheric temperature because a warmer atmosphere can hold more water vapor, shifting the absolute column-moisture threshold upward without requiring a greater degree of saturation for convection \citep{Kuo18}.

The coefficient $\alpha$ describes the sensitivity of the critical moisture threshold to thermal stratification  ($\mathrm{mm\ K^{-1}}$). With the sign convention in Eq.~\eqref{eq:qcrit}, a decrease in $\Delta$ during the moisture-recharge phase increases $q_{\mathrm{crit}}$ when $\alpha>0$, thereby suppressing convection and delaying its onset. Physically, this can be interpreted as representing a lower-tropospheric buoyancy barrier: if the decrease in $\Delta$ is primarily caused by warming of the lower-tropospheric environment relative to boundary-layer air, rising parcels become less buoyant. Then more moisture must accumulate to increase their moist static energy sufficiently to initiate deep convection.

The parameter $q_{\mathrm{hyst}}$ sets the amplitude of the hysteretic reduction in the moisture threshold (mm), allowing established convection to persist as atmospheric moisture decreases. The dimensionless variable $C$ represents the persistence of recent convective precipitation and evolves according to
\begin{equation}
\frac{\mathrm{d}C}{\mathrm{d}t}
=
\frac{P/P_{\max}-C}{\tau_C},
\label{eq:memory}
\end{equation}
where $\tau_C$ is the convective persistence timescale (days). Equation~\eqref{eq:memory} relaxes $C$ toward the normalized precipitation rate $P/P_{\max}$. $C\approx0$ represents a recently inactive convective state, whereas $C\approx1$
represents persistent precipitation near $P_{\max}$. For $C=0$, the critical moisture threshold is
\begin{equation}
q_{\mathrm{crit}}\big|_{C=0}
=
q_{\mathrm{crit},0}
-
\alpha
\left(
\Delta-\overline{\Delta}
\right),
\end{equation}
whereas for $C=1$, it is
\begin{equation}
q_{\mathrm{crit}}\big|_{C=1}
=
q_{\mathrm{crit},0}
-
\alpha
\left(
\Delta-\overline{\Delta}
\right)
-
q_{\mathrm{hyst}}.
\end{equation}
Convective persistence therefore lowers the critical moisture threshold by as much as $q_{\mathrm{hyst}}$, allowing established convection to continue at moisture levels below those required for its initial onset. Here, $C$ is a phenomenological representation of unresolved convective persistence rather than a directly diagnosed physical variable. Once convection is established, low-level moisture and temperature anomalies and cold-pool and mesoscale circulations can precondition the local environment and promote subsequent convection under conditions that would not initiate convection from a quiescent state \citep{Tompkins01a, Tompkins01b, Torri15, Colin19}. Without the persistence dependence in Eq.~(6), the discharge branch would lie at higher $q$ than the recharge branch for a given $\Delta$, because precipitation is activated only when $q>q_{\mathrm{crit}}$ in Eq.~(2). This would place the slow recharge branch in the lower left and the rapid discharge branch in the upper right, opposite to the branch ordering in the CRM (Figure~2(f)). By lowering $q_{\mathrm{crit}}$ after convection begins, Eq.~(6) introduces distinct onset and termination thresholds and is therefore necessary to reproduce both the direction and magnitude of the CRM hysteresis.

We next consider the deterministic skeleton of the COM by setting the stochastic-forcing amplitudes to zero. Because the precipitation rate satisfies
\begin{equation}
0 \leq P \leq P_{\max},
\label{eq:P_bounds}
\end{equation}
Eqs.~(3), (4), and (6) define the rectangular region
\begin{equation}
q_0-\tau_qP_{\max}
\leq q \leq q_0,
\label{eq:q_bounds}
\end{equation}
\begin{equation}
\Delta_0
\leq \Delta
\leq \Delta_0+\gamma\tau_\Delta P_{\max},
\label{eq:Delta_bounds}
\end{equation}
and
\begin{equation}
0 \leq C \leq 1.
\label{eq:C_bounds}
\end{equation}
Thus, all trajectories of the deterministic system remain bounded, precluding deterministic runaway solutions. These inequalities, however, do not constitute strict pathwise bounds for the full stochastic COM, because additive Gaussian white noise can drive $q$ and $\Delta$ beyond the deterministic invariant region. Nevertheless, the linear restoring terms in the $q$ and $\Delta$ equations, together with the bounded precipitation response, render the system mean-reverting and prevent finite-time runaway growth. Stochastic trajectories therefore fluctuate around the deterministic invariant region but are not confined to it.

The $q$--$\Delta$ deterministic subsystem is two-dimensional, the minimum phase-space dimension in which a smooth autonomous system can possess a nontrivial periodic orbit \citep{Strogatz18}. The prognostic variable $C$ extends the deterministic COM to three dimensions and represents a finite-memory hysteretic response to recent precipitation. In the stochastic system, periodic orbits are understood as properties of the deterministic skeleton, around which stochastic forcing produces irregular fluctuations.

\subsection{Linear Stability Analysis}

We perform a stability analysis using the governing equations in vector form \citep{Strogatz18}:
\begin{equation}
\mathbf{x}
=
\begin{pmatrix}
q\\
\Delta\\
C
\end{pmatrix},
\qquad
\frac{\mathrm{d}\mathbf{x}}{\mathrm{d}t}
=
\mathbf{F}(\mathbf{x})
=
\begin{pmatrix}
F_1\\
F_2\\
F_3
\end{pmatrix}
=
\begin{pmatrix}
\frac{q_0-q}{\tau_q}-P\\
\frac{\Delta_0-\Delta}{\tau_\Delta}+\gamma P\\
\frac{1}{\tau_C}
\left(
\frac{P}{P_{\max}}-C
\right)
\end{pmatrix}.
\label{eq:vector_system}
\end{equation}
We linearize the system about an equilibrium
\begin{equation}
\mathbf{x}^{*}
=
\begin{pmatrix}
q^{*}\\
\Delta^{*}\\
C^{*}
\end{pmatrix},
\qquad
\mathbf{F}(\mathbf{x}^{*})
=
\mathbf{0}.
\label{eq:equilibrium}
\end{equation}
From Eq.~(3), (4), and (6), ignoring the stochastic terms, the equilibrium conditions lead to
\begin{equation}
P^{*}
=
\frac{q_0-q^{*}}{\tau_q}
=
\frac{\Delta^{*}-\Delta_0}{\gamma\tau_\Delta}
=
P_{\max}C^{*}.
\label{eq:equilibrium_relations}
\end{equation}
Introducing the perturbation vector
\begin{equation}
\mathbf{x}'
=
\mathbf{x}-\mathbf{x}^{*}
=
\begin{pmatrix}
q'\\
\Delta'\\
C'
\end{pmatrix},
\label{eq:perturbation}
\end{equation}
the linearized system is
\begin{equation}
\frac{\mathrm{d}\mathbf{x}'}{\mathrm{d}t}
=
\mathbf{J}^{*}\mathbf{x}',
\label{eq:linearized_system}
\end{equation}
where the Jacobian is
\begin{equation}
\mathbf{J}^{*}
=
\left.
\begin{pmatrix}
\dfrac{\partial F_1}{\partial q}
&
\dfrac{\partial F_1}{\partial\Delta}
&
\dfrac{\partial F_1}{\partial C}
\\[8pt]
\dfrac{\partial F_2}{\partial q}
&
\dfrac{\partial F_2}{\partial\Delta}
&
\dfrac{\partial F_2}{\partial C}
\\[8pt]
\dfrac{\partial F_3}{\partial q}
&
\dfrac{\partial F_3}{\partial\Delta}
&
\dfrac{\partial F_3}{\partial C}
\end{pmatrix}
\right|_{\mathbf{x}=\mathbf{x}^{*}}.
\label{eq:jacobian_definition}
\end{equation}

\noindent Differentiating the precipitation relation in Eq.~(2) with respect to $q$ and evaluating the result at equilibrium, we define the equilibrium precipitation sensitivity as
\begin{equation}
G^{*}
\equiv
\left.
\frac{\partial P}{\partial q}
\right|_{\mathbf{x}=\mathbf{x}^{*}}
=
\frac{P_{\max}}{2q_{\mathrm{width}}}
\operatorname{sech}^{2}
\left(
X^{*}
\right),
\label{eq:Gstar}
\end{equation}
where
\begin{equation}
X^{*}
=
\frac{
q^{*}-q_{\mathrm{crit},0}
+\alpha
\left(
\Delta^{*}-\overline{\Delta}
\right)
+q_{\mathrm{hyst}}C^{*}
}{
q_{\mathrm{width}}
}.
\label{eq:Xstar}
\end{equation}
$G^{*}$ has units of $\mathrm{day^{-1}}$. The remaining precipitation derivatives are
\begin{equation}
\left.
\frac{\partial P}{\partial\Delta}
\right|_{\mathbf{x}=\mathbf{x}^{*}}
=
\alpha G^{*},
\qquad
\left.
\frac{\partial P}{\partial C}
\right|_{\mathbf{x}=\mathbf{x}^{*}}
=
q_{\mathrm{hyst}}G^{*}.
\label{eq:P_derivatives}
\end{equation}
The Jacobian evaluated at the equilibrium is 
\begin{equation}
\mathbf{J}^{*}
=
\begin{pmatrix}
-\dfrac{1}{\tau_q}-G^{*}
&
-\alpha G^{*}
&
-q_{\mathrm{hyst}}G^{*}
\\[10pt]
\gamma G^{*}
&
-\dfrac{1}{\tau_\Delta}
+\alpha\gamma G^{*}
&
\gamma q_{\mathrm{hyst}}G^{*}
\\[10pt]
\dfrac{G^{*}}{P_{\max}\tau_C}
&
\dfrac{\alpha G^{*}}{P_{\max}\tau_C}
&
\dfrac{q_{\mathrm{hyst}}G^{*}}{P_{\max}\tau_C}
-\dfrac{1}{\tau_C}
\end{pmatrix}.
\label{eq:jacobian}
\end{equation}

\noindent The characteristic polynomial of the Jacobian is
\begin{equation}
\det
\left(
\lambda\mathbf{I}-\mathbf{J}^{*}
\right)
=
\lambda^3
+a_1\lambda^2
+a_2\lambda
+a_3,
\label{eq:characteristic_polynomial}
\end{equation}
where
\begin{equation}
a_1
=
-\operatorname{tr}
\left(
\mathbf{J}^{*}
\right),
\end{equation}
\begin{equation}
a_2
=
\frac{1}{2}
\left\{
\left[
\operatorname{tr}
\left(
\mathbf{J}^{*}
\right)
\right]^2
-
\operatorname{tr}
\left[
\left(
\mathbf{J}^{*}
\right)^2
\right]
\right\},
\end{equation}
and
\begin{equation}
a_3
=
-\det
\left(
\mathbf{J}^{*}
\right).
\end{equation}

\noindent These coefficients are given as
\begin{equation}
a_1
=
\frac{1}{\tau_q}
+\frac{1}{\tau_\Delta}
+\frac{1}{\tau_C}
+
G^{*}
\left(
1-\alpha\gamma
-\frac{q_{\mathrm{hyst}}}{P_{\max}\tau_C}
\right),
\label{eq:a1}
\end{equation}
\begin{equation}
\begin{split}
a_2
={}&
\frac{1}{\tau_q\tau_\Delta}
+\frac{1}{\tau_q\tau_C}
+\frac{1}{\tau_\Delta\tau_C}
\\
&+
G^{*}
\left[
\frac{1}{\tau_C}
+\frac{1}{\tau_\Delta}
-\alpha\gamma
\left(
\frac{1}{\tau_C}
+\frac{1}{\tau_q}
\right)
\right.
\\
&\left.
\hspace{2.5cm}
-\frac{q_{\mathrm{hyst}}}{P_{\max}\tau_C}
\left(
\frac{1}{\tau_q}
+\frac{1}{\tau_\Delta}
\right)
\right],
\end{split}
\label{eq:a2}
\end{equation}
and
\begin{equation}
a_3
=
\frac{1}{\tau_C}
\left[
\frac{1}{\tau_q\tau_\Delta}
+
G^{*}
\left(
\frac{1}{\tau_\Delta}
-\frac{\alpha\gamma}{\tau_q}
-\frac{q_{\mathrm{hyst}}}
{P_{\max}\tau_q\tau_\Delta}
\right)
\right].
\label{eq:a3}
\end{equation}

The local behavior of perturbations is determined by the real and imaginary parts of the eigenvalues $\lambda_i$ ($i=1,2,3$) of $\mathbf{J}^{*}$ \citep{Strogatz18}. If all three eigenvalues are real and negative, perturbations decay monotonically and the equilibrium is a \textit{stable node}. If all three eigenvalues are real and positive, perturbations grow monotonically and the equilibrium is an \textit{unstable node}. If all eigenvalues are real but have mixed signs, the equilibrium is a \textit{saddle}. If one eigenvalue $\lambda_1$ is real while the remaining two form a complex-conjugate pair, $\lambda_{2,3}=\mu\pm i\omega$ with $\omega\neq0$, then $\mu=\operatorname{Re}(\lambda_{2,3})$ is the common real part of the pair and determines the exponential growth or decay rate of the oscillatory perturbations, while $\omega$ is their angular frequency. When $\lambda_1<0$ and $\mu<0$, the equilibrium is a \textit{stable focus}, corresponding to damped oscillations. When $\lambda_1>0$ and $\mu>0$, perturbations grow in every eigendirection and the equilibrium is an \textit{unstable focus}. When $\lambda_1$ and $\mu$ have opposite signs, some perturbations grow while others decay, and the equilibrium is a \textit{saddle-focus}.

According to the Routh--Hurwitz criterion, the equilibrium is linearly stable if and only if
\begin{equation}
a_1>0,
\qquad
a_2>0,
\qquad
a_3>0,
\qquad
H \equiv a_1a_2-a_3>0.
\label{eq:routh_hurwitz}
\end{equation}
Provided that $a_1$, $a_2$, and $a_3$ remain positive, the oscillatory stability transition is determined by
\begin{equation}
\begin{cases}
H>0,
& \text{linearly stable},\\
H=0,
& \text{oscillatory stability boundary},\\
H<0,
& \text{oscillatory instability}.
\end{cases}
\label{eq:stability_transition}
\end{equation}

At the oscillatory stability boundary, $a_3=a_1a_2$, and the characteristic polynomial factorizes as
\begin{equation}
\lambda^3
+a_1\lambda^2
+a_2\lambda
+a_1a_2
=
\left(
\lambda+a_1
\right)
\left(
\lambda^2+a_2
\right).
\label{eq:factorization}
\end{equation}
The corresponding eigenvalues are
\begin{equation}
\lambda_1
=
-a_1,
\qquad
\lambda_{2,3}
=
\pm i\sqrt{a_2}.
\label{eq:hopf_eigenvalues}
\end{equation}
Accordingly, $H=0$, subject to $a_1,a_2,a_3>0$, defines an oscillatory linear-stability boundary and hence a candidate Hopf bifurcation.

We obtain the COM parameters required for the stability analysis and numerical simulations from the CRM time series of $q$ and $\Delta$ at 305 and 325~K. We set $P_{\max}=200$, based on the maximum precipitation repeatedly attained during the episodic 325~K simulation, and use the same value for the 305~K case, treating $P_{\max}$ as a common asymptotic scale of the precipitation closure rather than as a case-specific sample maximum. From the hysteretic loop in Figure~2(f), we estimate $q_{\mathrm{hyst}}=17.40$ at 325~K as the difference between the maximum and minimum moisture values, $q_{\max}-q_{\min}$. We set $q_{\mathrm{hyst}}=0$ at 305~K because Figure~2(e) shows no clear hysteresis. We set $q_{\mathrm{width}}=0.40$ for both cases, consistent with the moisture--precipitation transition width reported by \citet{Kuo18}, who found no systematic change in the sharpness of this transition with warming. Given these numbers, we estimate $\tau_q$, $\tau_\Delta$, $\tau_C$, $\alpha$, $\gamma$, and $q_{crit,0}$ by regression against the CRM time series. For the 305~K case, $\tau_C$ is not estimated because no clear hysteresis is present. Finally, we estimate the stochastic amplitudes $\sigma_q$ and $\sigma_T$ from the corresponding regression residuals of the 305~K simulation \citep[e.g.,][]{Han26a} and apply the same values to the 325~K case. This is a simplifying assumption, as we do not treat changes in stochastic forcing as an independent control on the regime transition. Because the noise is additive, $\sigma_q$ and $\sigma_T$ do not enter the Jacobian or the linear stability analysis; however, they influence the variability around the deterministic trajectory and therefore affect the spread and visual realism of the simulated phase diagrams.

Table~1 summarizes all parameter values, obtained eigenvalues, and the classified equilibrium regime, which shows that the 305~K case indeed corresponds to damped oscillations while the 325~K case corresponds to unstable oscillations, as suggested by the time series in Figure~2.

\begin{table}[!htb]
\caption{COM parameter values for the 305 and 325~K simulations.}
\label{tab:com_parameters}
\centering
\begin{tabular}{lcc}
\hline
Parameter & 305~K & 325~K \\
\hline
$\tau_q$ (day)          & 1.50 & 2.05 \\
$\tau_\Delta$ (day)     & 0.45 & 8.79 \\
$\tau_C$ (day)         & - & 0.01 \\
$\alpha$ (mm K$^{-1}$)         & 2.04 & 0.01 \\
$\gamma$ (K mm$^{-1}$)          & 0.27 & 0.29 \\
$q_{\mathrm{hyst}}$ (mm) & 0 & 17.40 \\
$P_{\max}$ (mm day$^{-1}$)        & 200 & 200 \\
$q_{\mathrm{width}}$ (mm) & 0.40 & 0.40 \\
$q_{\mathrm{crit},0}$ (mm) & 72.81 & 374.99 \\
$\sigma_q$ (mm day$^{-0.5}$)     & 0.25 & 0.25 \\
$\sigma_T$ (K day$^{-0.5}$)     & 0.10 & 0.10 \\
$\lambda_1$ (day$^{-1}$)      & -81.89 & -0.11 \\
$\lambda_{2,3}$ (day$^{-1}$)      & $-5.80 \pm 2.65i$ & $2.49 \pm 30.08i$ \\
Regime       & Stable focus (damped oscillation) & Saddle focus (unstable oscillation) \\
\hline
\end{tabular}
\end{table}

\begin{figure}[htbp]
  \centering
  \includegraphics[width=0.70\textwidth]{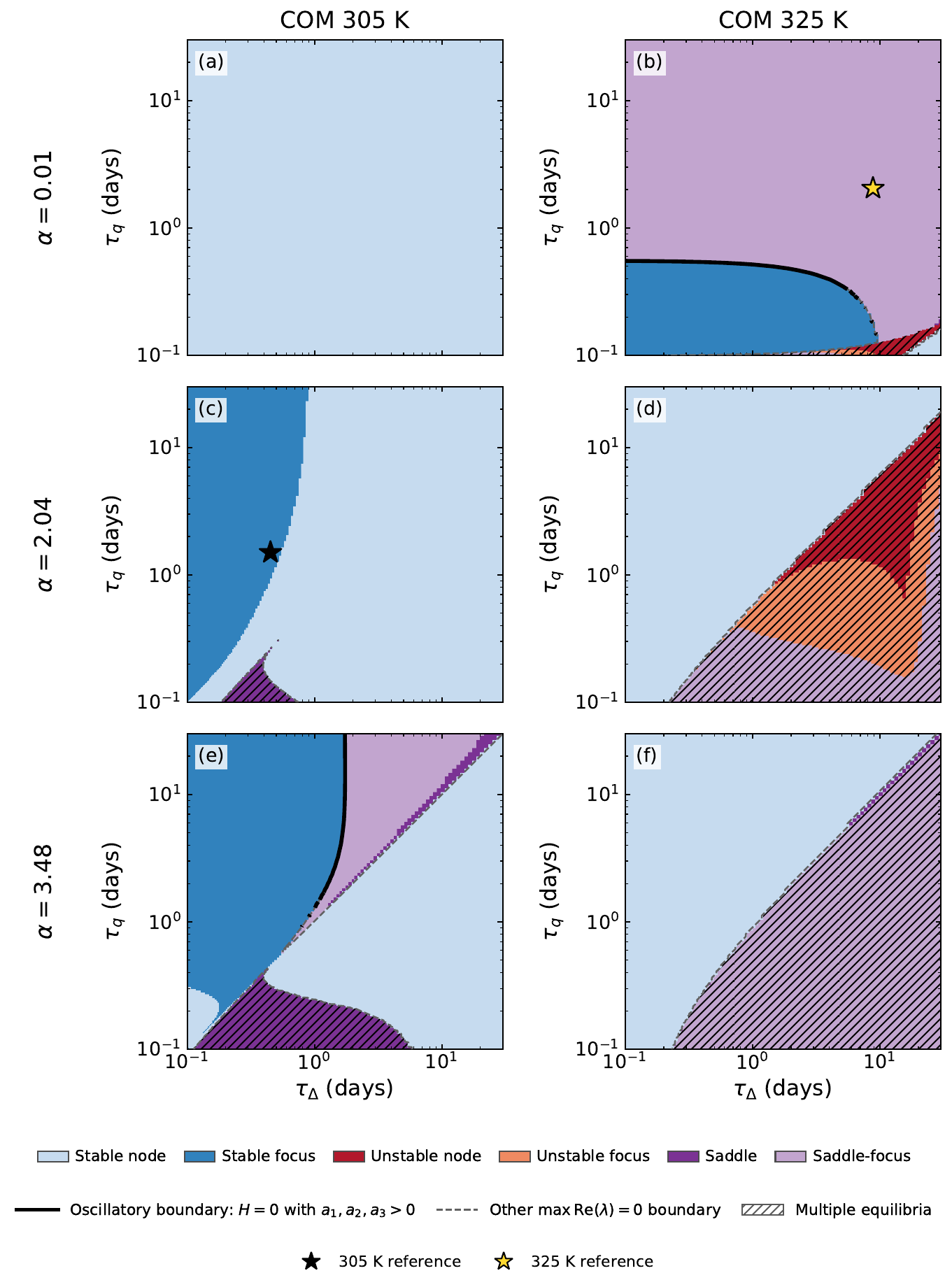}
  \caption{
  Linear stability regimes in $(\tau_\Delta,\tau_q)$ parameter space for the 305~K (panels (a,c,e)) and 325~K (panels (b,d,f)) cases. For each case, all parameters are fixed at their CRM-fitted reference values given in Table~1, except for $\tau_\Delta$ and $\tau_q$, which vary along the two axes, and $\alpha$, which is set to 0.01, 2.04, and 3.48 in the first, second, and third rows, respectively. Solid black contours denote candidate Hopf boundaries. Dashed gray contours denote other neutral-stability boundaries where $\max\mathrm{Re}(\lambda)=0$, but the complete Routh--Hurwitz conditions for an oscillatory boundary are not satisfied. The black star in panel (c) marks the reference 305~K simulation at $(\tau_\Delta,\tau_q)=(0.45,1.50)$~days, which lies in the stable-focus (damped-oscillation) regime. The yellow star in panel (b) marks the reference 325~K simulation at $(\tau_\Delta,\tau_q)=(8.79,2.05)$~days, which lies in the saddle-focus (unstable-oscillation) regime.
  }
\end{figure}

Figure~3 visualizes the linear stability regimes in $(\tau_\Delta,\tau_q)$ parameter space. We focus on these two timescales and the sensitivity coefficient $\alpha$ because they directly control the pace and strength of the moisture–stratification feedback. Again, $\tau_q$ governs moisture recharge toward $q_0$ through surface evaporation, $\tau_\Delta$ governs the radiative and dynamical adjustment of stratification toward $\Delta_0$, and $\alpha$ determines how strongly stratification modifies the critical moisture threshold for convection. The first, second, and third rows show $\alpha=0.01$, 2.04, and 3.48~$\mathrm{mm\ K^{-1}}$, respectively. The values $\alpha=0.01$ and 2.04 are fitted to the 325 and 305~K cases, respectively, whereas $\alpha=3.48$ illustrates how stronger moisture–stratification sensitivity can produce an unstable oscillatory regime within the explored parameter space. The left and right columns correspond to the 305 and 325~K cases, respectively, with all remaining parameters fixed at their case-specific values given in Table~1. For the 305~K case, $\alpha=0.01$ produces a stable node throughout the displayed $(\tau_\Delta,\tau_q)$ parameter space (Figure~3(a)). Increasing $\alpha$ to 2.04 produces a stable-focus regime, corresponding to damped oscillations, within which the reference 305~K COM simulation lies; however, no saddle-focus (self-sustained oscillation) regime occurs over the displayed parameter range (Figure~3(c)). When $\alpha$ is further increased to 3.48, an oscillatory stability boundary—a candidate Hopf boundary—emerges between the stable-focus and saddle-focus regimes (Figure~3(e)); increasing $\tau_\Delta$ and $\tau_q$ together can move the system across the stability boundary from a damped to an unstable, self-sustained oscillator, whereas decreasing both timescales can restore damped oscillatory behavior.

For the 325~K case, a saddle-focus regime is already present at $\alpha=0.01$, and the reference parameter combination lies within this regime (Figure~3(b)), consistent with an unstable oscillator. Crossing the candidate Hopf boundary by reducing $\tau_{\Delta}$ and  $\tau_{q}$ leads from the saddle-focus regime to a stable-focus regime. As $\alpha$ increases to 2.04 and 3.48, the unique equilibrium gives way to multiple equilibria over parts of the $(\tau_\Delta,\tau_q)$ parameter space (Figures~3(d,f)). Consequently, the realized state in this parameter space may depend on the stability of the coexisting equilibria and on the initial conditions. However, this multiple-equilibrium regime lies outside the calibrated range and is therefore not investigated further here.

\section{Coupled Oscillator Model Numerical Results}

We perform numerical COM simulations using the parameter sets listed in Table~1, which were fitted separately to the corresponding 305 and 325~K CRM simulations. The stochastic COM is integrated using the Euler--Maruyama scheme with a timestep of 0.01~day. Each simulation is run for 200~days, and the final 20~days are presented. Figure~4 shows the resulting time series of $\Delta$, $q$, and $P$, together with the corresponding phase diagrams, in the same format as Figure~2. These results test whether the reduced equations reproduce the temporal and phase-space characteristics that cannot be inferred from linear stability analysis alone.

As a damped oscillator, the COM 305~K case exhibits noisy and anticorrelated behavior between $\Delta$ and $q$ (Figure~4(c)), which translates to quasi-steady precipitation with  the precipitation index $\eta=0.32<1$ (Figure~4(a)), manifesting no obvious closed orbit (Figure~4(e)). For the 325~K case, there appear repeated regular cycles of $\Delta$ and $q$, still anticorrelated (Figure~4(d)), which give rise to episodic convection when $q$ is discharged (Figure~4(b)); the precipitation index is $\eta=5.96>1$. The $\Delta$--$q$ phase diagram exhibits a limit cycle with hysteresis, where a more packed slow recharge branch is located toward the upper right and a sparser fast discharge branch is located toward the lower left (Figure~4(f)). Overall, this illustrates that our conceptual model with $q$ and $\Delta$, extended by convective persistence (variable $C$), can reproduce the quasi-steady convection as stable, noise-driven oscillations and episodic convection as unstable, self-sustained oscillations with the hysteretic behavior as in the CRM (compare Figures~2 and 4).

\begin{figure}[htbp]
  \centering
  \includegraphics[width=1.0\textwidth]{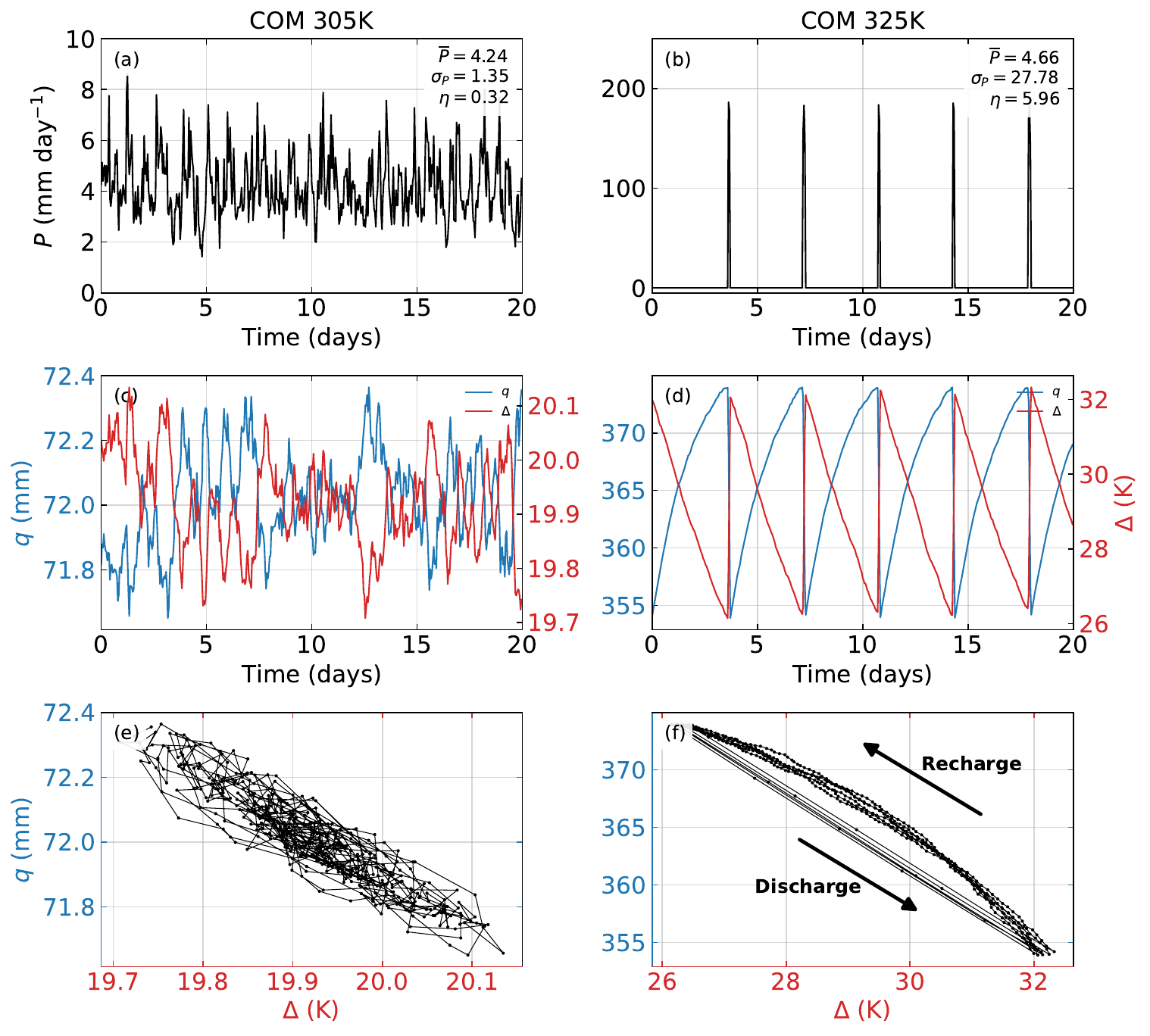}
  \caption{
  As in Figure~2, but generated from COM simulations using the fitted CRM parameter values listed in Table~1. 
  }
\end{figure}

Figure~5 evaluates whether the COM reproduces the statistical distributions of the CRM variables. The left and right columns show the 305 and 325~K cases, respectively, while the rows show column water vapor $q$, $\Delta$, and $P$. At 305~K, both CRM and COM exhibit narrow, approximately unimodal distributions of $q$, $\Delta$, and $P$. The close agreement indicates that the COM captures the weakly varying, quasi-steady convective regime, although the COM overestimates the upper tail of the precipitation PDF. At 325~K, the distributions become much broader and strongly non-Gaussian. Both models exhibit a wide range of $q$ and $\Delta$, frequent periods of nearly zero precipitation, and rare events approaching $200~\mathrm{mm\ day^{-1}}$, consistent with intense convective discharge followed by dry-spells. The COM therefore reproduces the emergence of episodic convection and its large precipitation variability, although it produces a sharper accumulation near the bottom and upper end of the $q$ distribution (Figure~5(b)) and more pronounced peaks in the near-zero and intense-precipitation states without much at intermediate values (Figure~5(f)).

\begin{figure}[htbp]
  \centering
  \includegraphics[width=1.0\textwidth]{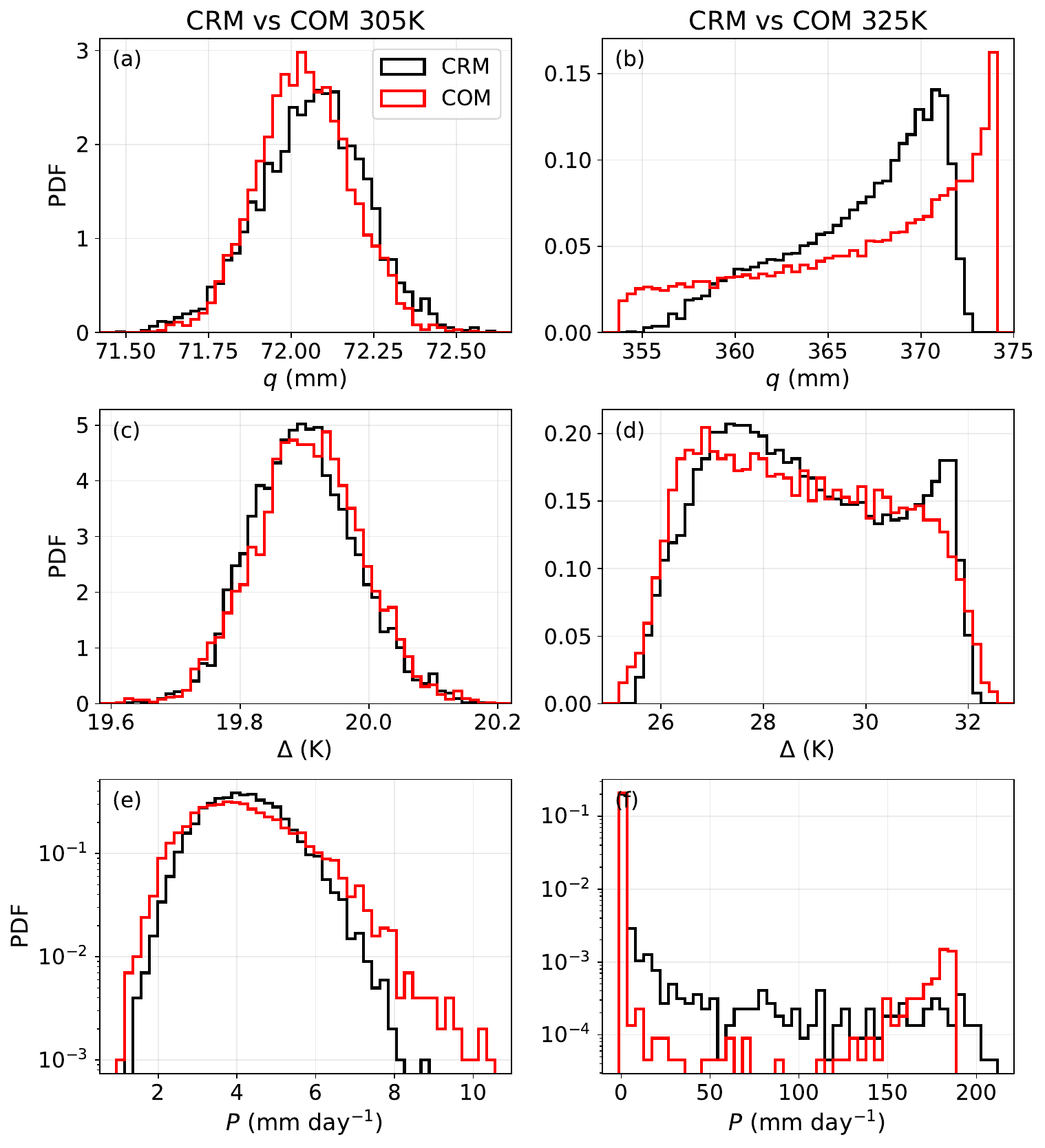}
  \caption{
  Comparison of the probability density functions of $q$ (panels a,b), $\Delta$ (panels c,d), and $P$ (panels e,f) between the CRM and COM simulations. The left column (panels a,c,e) shows the 305~K case, and the right column (panels b,d,f) shows the 325~K case. Black and red curves denote the CRM and COM results, respectively.
  }
\end{figure}

We further compare power spectra between the COM and the CRM (Figure~6). At 305~K, the spectra of $q$ shows a good match between the models (Figure~6(a)), whereas for $\Delta$ the low frequency component is overestimated and high frequency component is underestimated in the COM compared to the CRM (Figure~6(c)). For $P$, the low-frequency part (longer than one day) is excessive for COM (Figure~6(e)). The 325~K spectra contain a strong fundamental peak at a period of approximately 3--4 days and a sequence of higher-frequency harmonics, while over-representing the low frequency component (Figure~6(f)). These peaks also occur at corresponding periods in $q$ and $\Delta$ (Figure~6(b,d)); the COM closely reproduces both the dominant period and the harmonic structure of the CRM, demonstrating that it captures not only the enhanced variability of the episodic regime but also the recurrence timescale and strongly non-sinusoidal character of the convective events.

\begin{figure}[htbp]
  \centering
  \includegraphics[width=1.0\textwidth]{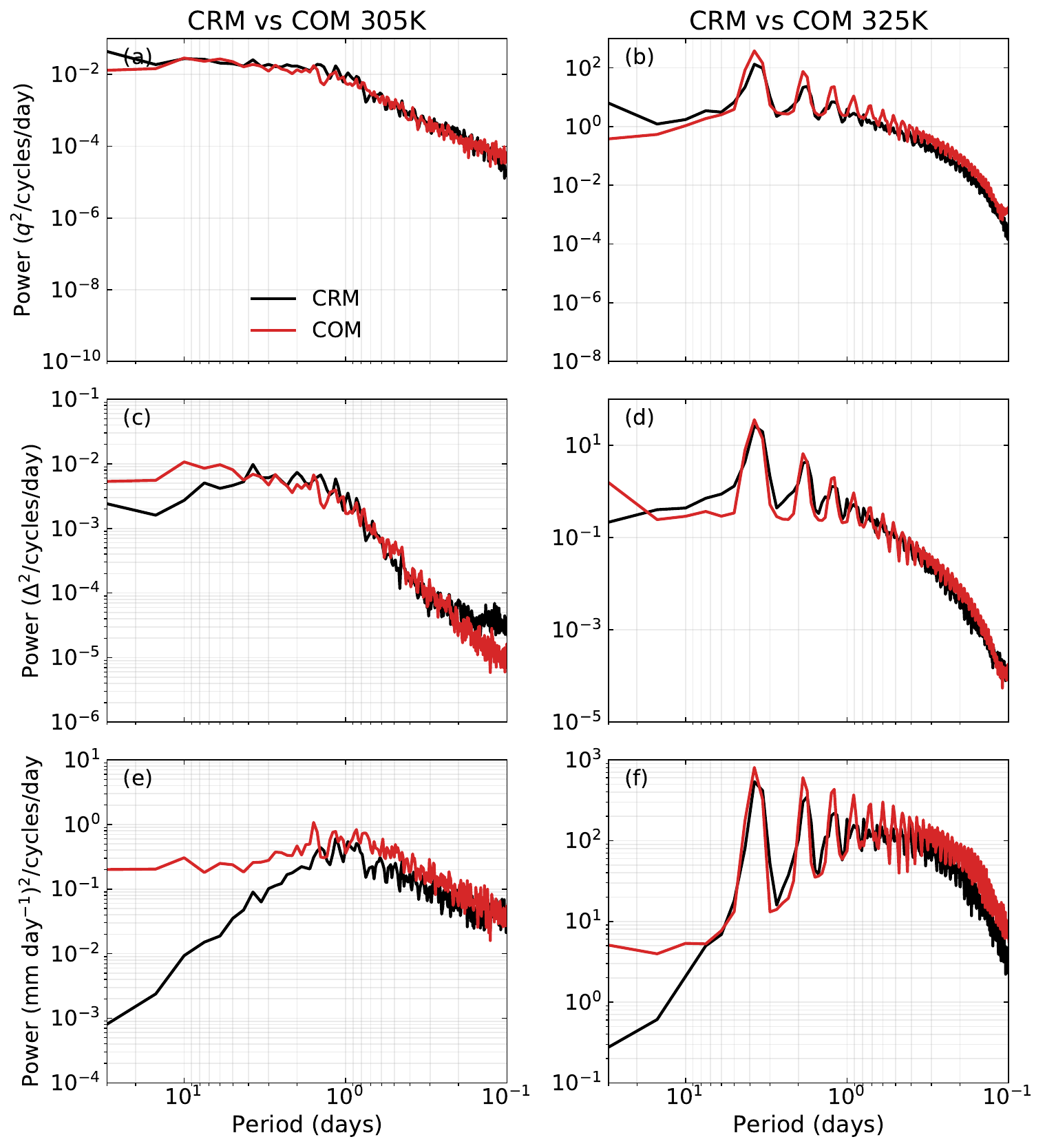}
  \caption{
  Comparison of the power spectra of $q$ (panels a,b), $\Delta$ (panels c,d), and $P$ (panels e,f) between the CRM and COM simulations. The left column (panels a,c,e) shows the 305~K case, and the right column (panels b,d,f) shows the 325~K case. Black and red curves denote the CRM and COM results, respectively.
  }
\end{figure}

Taken together, Figures~4--6 demonstrate that the highly-reduced COM simulates quasi-steady convection as a stable, noise-driven oscillation and episodic convection as self-sustained oscillations, closely reproducing the results of the CRM, including the coupled $\Delta$--$q$ evolution, hysteresis, statistical distributions, and power spectra. Although quantitative discrepancies remain, the overall agreement shows that moisture, thermal stratification, and convective persistence provide an adequate minimum framework for capturing the principal dynamics of both convective regimes and the transitions between them.

\section{Discussions and Conclusion}

We developed a highly reduced-order COM that couples column-integrated moisture, thermal stratification, and convective persistence to simulate convective precipitation in different regimes. Despite its simplicity, the COM reproduces the principal characteristics of the CRM simulations: quasi-steady convection expressed as noisy and damped oscillations, and episodic convection expressed as self-sustained oscillations with hysteresis. These results suggest that the transition can be understood as a change in the stability of a coupled moisture--thermal system rather than solely as a consequence of a particular radiative or dynamical mechanism.

The convective-regime transition occurs when the system can no longer damp coupled moisture--thermal perturbations before sufficient moisture accumulates to activate the nonlinear precipitation feedback. At 305~K, the relatively short relaxation timescales, $(\tau_\Delta,\tau_q)=(0.45,1.50)$~days, rapidly return $\Delta$ and $q$ toward stable equilibrium. Meanwhile, the relatively large $\alpha=2.04$ makes the critical moisture threshold strongly sensitive to thermal stratification; when $\Delta$ is elevated, $q_{\mathrm{crit}}$ is lowered (Eq.~(7)), allowing precipitation to activate before substantial moisture accumulates. This early moisture discharge helps damp perturbations, leaving stochastic forcing to maintain only weak fluctuations. At 325~K, the longer relaxation timescales, $(\tau_\Delta,\tau_q)=(8.79,2.05)$~days, allow moisture and thermal anomalies to persist. Meanwhile, the small $\alpha=0.01$ makes $q_{\mathrm{crit}}$ nearly insensitive to changes in $\Delta$. During the dry phase, $q$ accumulates while $\Delta$ decreases until $q$ exceeds the relatively high baseline threshold $q_{\mathrm{crit},0}$, causing precipitation to increase sharply. Precipitation then rapidly depletes moisture and increases $\Delta$, while the state-dependent convective-persistence term lowers $q_{\mathrm{crit}}$, allowing convection to persist during discharge and producing hysteresis. Once moisture is sufficiently depleted, convection terminates and the recharge phase begins again. The combination of slow moisture recharge, slow thermal recovery, and convective persistence thus transforms damped, noise-driven fluctuations into a self-sustained recharge--discharge cycle.

The increase in $\tau_q$ is consistent with the rapidly increasing atmospheric moisture reservoir under warming, following Calusis-Clapeyron relation, while evaporation increases more slowly because it remains constrained by the atmospheric energy budget \citep[e.g.,][]{Schneider10, Gorman12, Liu24}. Recharge from the moisture deficit produced by a convective event therefore takes longer in the warmer atmosphere. The substantially larger $\tau_\Delta$ indicates that convectively generated stratification also relaxes much more slowly in the warmer simulations. This slow thermal recovery allows convective inhibition to persist while moisture recharges, separating the slow recharge phase from the rapid convective discharge \citep{Seeley21, Spaulding24, Yang24}. Because thermal stratification can be modified by LTRH and vertical contrasts in radiative cooling \citep{Seeley21, Song24}, $\tau_\Delta$ should be interpreted as an effective bulk timescale representing the combined effects of thermal relaxation processes rather than as a purely radiative timescale.

The fitted value of $\alpha$ decreases in the warmer simulation, indicating that the critical moisture threshold becomes substantially less sensitive to thermal-stratification anomalies. This behavior may be related to the documented dependence of the moisture threshold on the atmospheric thermal environment \citep{Kuo18,Ahmed18,Neelin22}, although the direct physical interpretation of $\alpha$ is not straightforward. The mechanisms underlying this apparent warming dependence should therefore be investigated further. The parameter $\gamma$ measures the increase in thermal stratification per unit precipitation and therefore represents convective stabilization. Its fitted values differ by only approximately 10\% between the two cases, suggesting that the efficiency with which convection stabilizes the atmospheric column is not the principal cause of the regime transition. This similarity reflects the weak temperature dependence of the latent heat released per unit mass of condensed water, although the vertical distributions of condensational heating and evaporative cooling may still change with warming.

Our selection of $q$ and $\Delta$ as prognostic variables represents one particular reduction of the atmospheric thermodynamic state. \citet{Kuo18} showed that the critical column-water-vapor threshold depends systematically on bulk tropospheric temperature: convective onset occurs at higher column water vapor, but lower column relative humidity, as the troposphere warms. They also found vertically coherent temperature variations, supporting the use of a bulk tropospheric temperature measure in convective-transition statistics. In the COM, this mean-state temperature dependence is incorporated implicitly through the case-specific value of $q_{\mathrm{crit},0}$, whereas $\Delta$ represents variations in vertical thermal stratification about that mean state. Conversely, \citet{Quan26} emphasized the vertical moisture gradient rather than bulk moisture content, showing that a stronger moisture gradient enhances wave-induced moistening and is particularly effective at destabilizing convectively coupled waves. Our model follows \citet{Stechmann11} in retaining column-integrated moisture as the moisture variable. Column moisture and its vertical gradient generally covary, but they are not interchangeable: columns with the same $q$ can have different vertical moisture distributions, buoyancy profiles, and entrainment characteristics. A natural extension of the COM would therefore divide moisture into lower- and upper-tropospheric reservoirs or introduce an explicit vertical moisture-gradient variable. This would clarify when column moisture provides an adequate proxy and when vertical moisture structure supplies an independent control on the convective regime.

The occurrence of episodic convection on Titan, a satellite of Saturn, further supports a framework that is not tied to one radiative mechanism. Episodic convection appears in simulations of Titan, where methane rather than water is the condensing species \citep{Lora15,Lora24,Spaulding24}. Observations reveal rare but intense methane storms, separated by several Earth years, that are sufficiently extensive as to observably modify Titan's surface \citep{Turtle11}. Titan is particularly instructive because it lacks the LTRH structure \citep[e.g.,][]{Tomasko08} invoked for hothouse convection on Earth. Moreover, Titan simulations produce quasi-steady precipitation in an idealized globally wet aquaplanet but episodic precipitation in a more realistic configuration with a dry equator, despite using the same radiative treatment \citep{Lora24}. This contrast implicates moisture availability and recharge through surface evaporation—which is regulated by near-surface winds, atmospheric stability, and turbulent surface–atmosphere exchange \citep[e.g.,][]{Han&Lora25}—as controls on the convective regime that operate independently of the particular radiative mechanisms.

Several limitations motivate future work. The COM is fitted to domain-mean output from two CRM simulations and does not explicitly represent spatial convective organization, moisture convergence, gravity-wave propagation, or cloud-radiative feedbacks. Its idealized threshold precipitation closure also contributes to discrepancies in the simulated precipitation distributions and spectra. Because the COM contains three prognostic variables, its nonlinear behavior beyond the saddle-focus may also include chaotic cycles \citep[e.g.,][]{Lorenz63}, as \citet{Sevellec14} demonstrated in a three-variable model of the Atlantic meridional overturning circulation. However, exploring such behavior is beyond the scope of this study. Analytically, future work should identify relevant nondimensional combinations of the governing timescales and feedback strengths, including $\tau_\Delta/\tau_q$, $\alpha\gamma$, $q_{\mathrm{hyst}}/q_{\mathrm{width}}$, and $\tau_C/\tau_q$. Additionally, the pronounced separation between the slow moisture-recharge and thermal-recovery phases and the rapid convective-discharge phase suggests that the episodic regime may be analyzed as a slow--fast dynamical system \citep[e.g.,][]{Strogatz18, Kuehn15}. Slow--fast analysis could approximate the oscillation period from the time spent along the slow recharge branch, determine the amplitude from the states at which rapid transitions occur, and quantify recharge--discharge asymmetry from the relative durations of the slow and fast phases. Testing these predictions across wider CRM, global climate simulations, and planetary parameter ranges will determine whether a common stability criterion can explain the onset of episodic convection.

In summary, the COM connects strong convection associated with a nonlinear moisture threshold to episodic convection characterized by recurrent temporal discharge. Quasi-steady convection behaves as a stable oscillator maintained by stochastic forcing, whereas episodic convection behaves as an unstable, self-sustained oscillator with hysteresis. The transition occurs when the combined effects of moisture recharge, thermal recovery, thermal sensitivity, and convective persistence destabilize the equilibrium and establish a repeating recharge--discharge cycle. The COM therefore provides a compact framework for isolating the essential mechanisms governing transitions between convective regimes across different climates and atmospheric configurations. Future work should test its predictive capability by diagnosing COM parameters across a wider range of simulations and evaluating whether the resulting stability criteria can anticipate when regime transitions occur and predict the frequency and intensity of episodic precipitation.

\acknowledgments
The authors have no conflicts of interest to disclose. S.H. thanks Yi-Hung Kuo for introducing him to the concept of strong convection and David Neelin for insightful discussions on bifurcation theory. These exchanges motivated the present work, which links strong and episodic convection within a dynamical-systems framework. S.H. also acknowledges the use of ChatGPT 5.6 for grammatical assistance in preparing the manuscript.  B. F. is supported by Yale's Skinner Postdoctoral fellowship. J. V. was supported by the ARISE project (ANR-18-CE01-0012) and the ``Tropico'' project funded by the LEFE program of the Institut des Sciences de l'Univers (INSU). A. V. F. was supported by NASA (80NSSC21K0558) and DOE (DE-SC0023134). 


%
%
\datastatement

The CRM data generated by \citet{Seeley21} and \citet{Song24} are archived on Zenodo as \citet{Seeley21Data} and \citet{Song23Data}, respectively. The COM code developed and used in this study will also be archived on Zenodo upon publication.


%






%




\appendix[A]
\appendixtitle{Additional Figures Using Simulation Cases from \citet{Song24}}

\begin{figure}[htbp]
  \centering
  \includegraphics[width=0.75\textwidth]{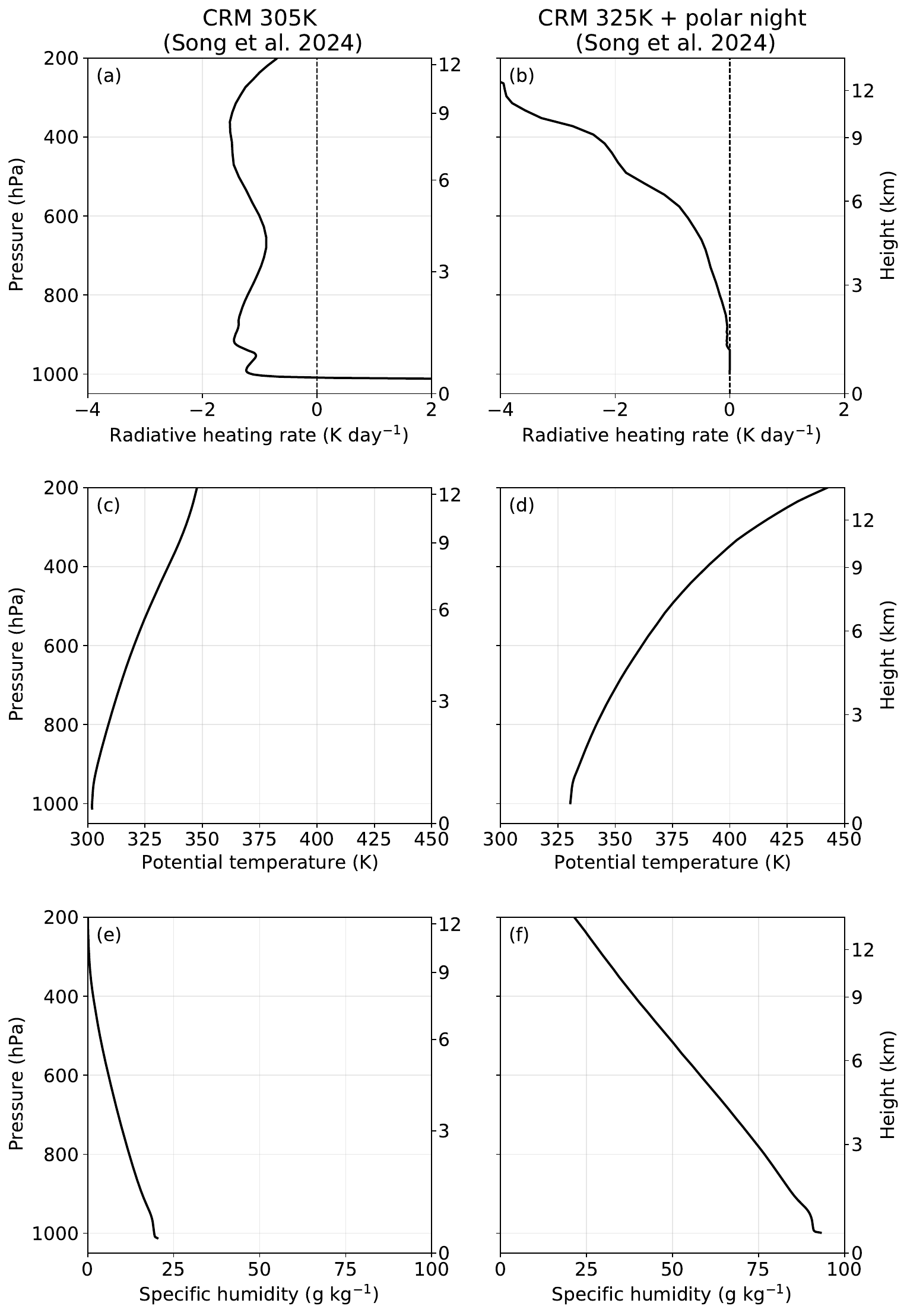}
  \caption{Similar to Figure~1, but using the CRM output using the 305~K and 325~K + polar night setup of \citet{Song24}.}
\end{figure}

\begin{figure}[htbp]
  \centering
  \includegraphics[width=1.0\textwidth]{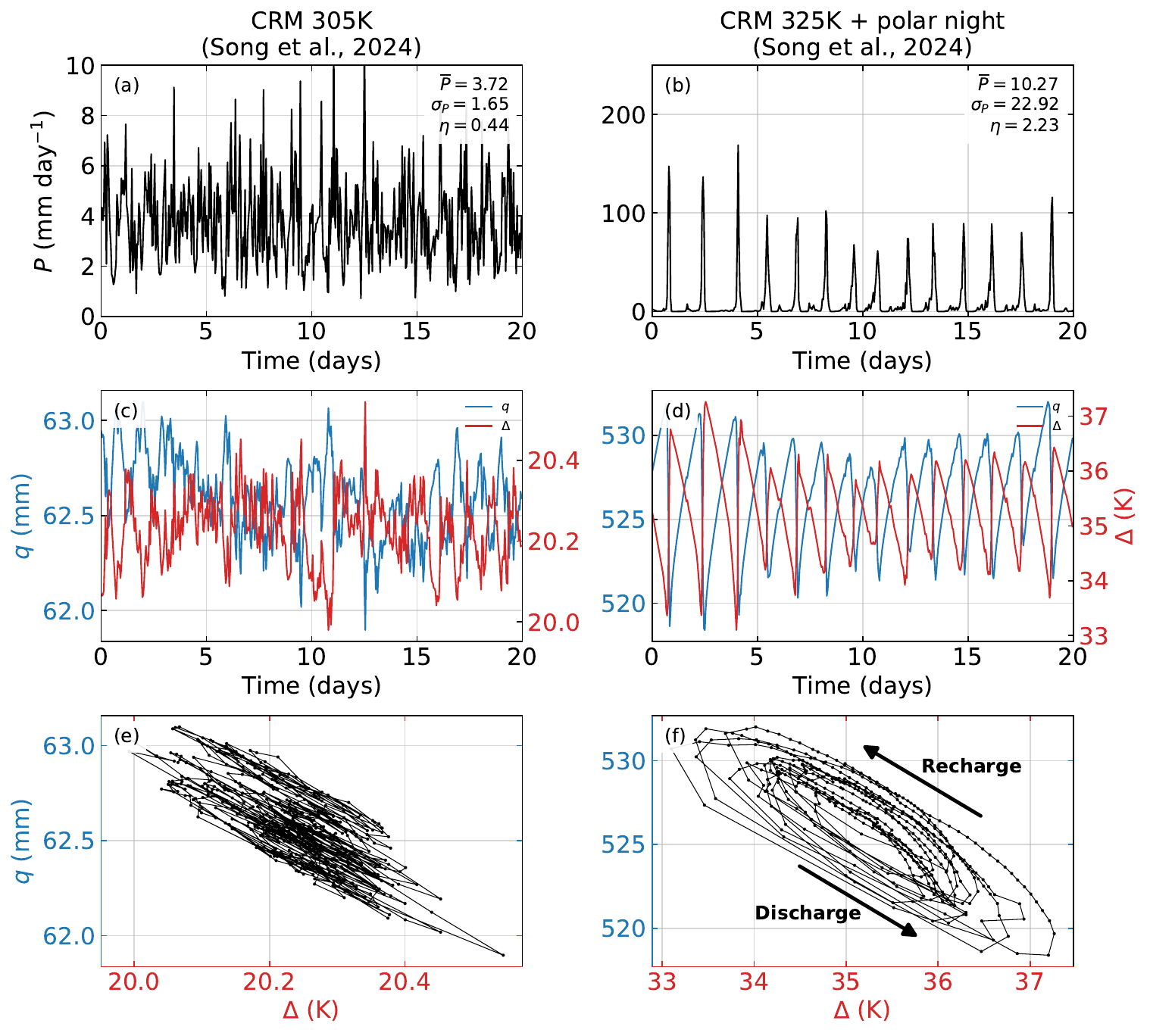}
  \caption{
  Similar to Figure~2, but using the CRM output using the 305~K and 325~K + polar night setup of \citet{Song24}.
  }
\end{figure}



\bibliographystyle{ametsocV6}
\bibliography{COM}

\end{document}